\PassOptionsToPackage{usenames,dvipsnames}{xcolor}

\documentclass[pdflatex,sn-nature,iicol]{sn-jnl}

\usepackage{graphicx}%
\usepackage{amsmath,amssymb,amsfonts}%
\usepackage{amsthm}%
\usepackage{mathrsfs}%
\usepackage[title]{appendix}%
\usepackage{xcolor}%
\usepackage{textcomp}%
\usepackage{manyfoot}%
\usepackage{booktabs}%
\usepackage{algorithm}%
\usepackage{algorithmicx}%
\usepackage{algpseudocode}%
\usepackage{listings}%
\usepackage{caption}
\usepackage{enumitem}
\usepackage[many]{tcolorbox}
\usepackage{setspace}
\usepackage{titlesec}
\usepackage{multicol}
\usepackage{multirow}%

\usepackage{helvet}
\renewcommand\familydefault\sfdefault
\titleformat{\section}
  {\normalfont\sffamily\large\bfseries}
  {\thesection}{1em}{}

\begin{document}

\title[Article Title]{\bf
    Robustness tests for biomedical foundation models should tailor to specifications \vspace{1em}
}

\author[1,2,*]{\fnm{R. Patrick} \sur{Xian}}

\author[3]{\fnm{Noah R.} \sur{Baker}}

\author[4]{\fnm{Tom} \sur{David}}

\author[1,5]{\fnm{Qiming} \sur{Cui}}

\author[6]{\fnm{A. Jay} \sur{Holmgren}}

\author[7]{\fnm{Stefan} \sur{Bauer}}

\author[8]{\fnm{Madhumita} \sur{Sushil}}

\author[1,2,9,*]{\fnm{Reza} \sur{Abbasi-Asl}}

\affil[1]{\orgdiv{Department of Neurology}, \orgname{University of California, San Francisco}, \orgaddress{\street{1651 4th Street}, \city{San Francisco}, \state{CA} \postcode{94158}, \country{USA}}}

\affil[2]{\orgdiv{Weill Institute for Neurosciences}, \orgname{University of California, San Francisco}, \orgaddress{\street{1651 4th Street}, \city{San Francisco}, \state{CA} \postcode{94158}, \country{USA}}}

\affil[3]{\orgdiv{Biological and Medical Informatics Graduate Program}, \orgname{University of California, San Francisco}, \orgaddress{\street{550 16th Street, 3rd Floor}, \city{San Francisco}, \state{CA} \postcode{94158}, \country{USA}}}

\affil[4]{\orgdiv{PRISM Eval}, \orgaddress{10 Rue de Penthièvre, \postcode{75008} \city{Paris}, \country{France}}}

\affil[5]{\orgdiv{Department of Bioengineering}, \orgname{University of California, Berkeley}, \orgaddress{\street{306 Stanley Hall, University Drive}, \city{Berkeley}, \state{CA} \postcode{94720}, \country{USA}}}

\affil[6]{\orgdiv{Division of Clinical Informatics and Digital Transformation}, \orgname{University of California, San Francisco}, \orgaddress{\street{10 Koret Way}, \city{San Francisco}, \state{CA} \postcode{94117}, \country{USA}}}

\affil[7]{School of Computation, Information and Technology, Technical University of Munich \& Helmholtz AI, \orgaddress{\street{Friedrich-Ludwig-Bauer-Strasse 5}, \postcode{85748} \city{Garching bei München}, \country{Germany}}}

\affil[8]{\orgdiv{Bakar Computational Health Sciences Institute}, \orgname{University of California, San Francisco}, \orgaddress{\street{490 Illinois Street}, \city{San Francisco}, \state{CA} \postcode{94158}, \country{USA}}}

\affil[9]{\orgdiv{Department of Bioengineering and Therapeutic Sciences}, \orgname{University of California, San Francisco}, \orgaddress{\street{1700 4th Street}, \city{San Francisco}, \state{CA} \postcode{94143}, \country{USA}}}

\affil[*]{Corresponding authors: \texttt{xrpatrick\,@\,gmail.com, reza.abbasiasl\,@\,ucsf.edu}}

\onecolumn
\maketitle
\setstretch{1.2}

\section*{Abstract}
\vspace{-0.5em}
\noindent\textbf{The rise of biomedical foundation models creates new hurdles in model testing and authorization, given their broad capabilities and susceptibility to complex distribution shifts. We suggest tailoring robustness tests according to task-dependent priorities and propose to integrate granular notions of robustness in a predefined specification to guide implementation. Our approach facilitates the standardization of robustness assessments in the model lifecycle and connects abstract AI regulatory frameworks with concrete testing procedures.
}

\setstretch{1.55}
The growing presence of biomedical foundation models (BFMs), including large language models (LLMs), vision-language models (VLMs), and others, trained using biomedical or de-identified healthcare data, suggests they will eventually become integral to healthcare automation. Discussions on the risks of deploying algorithmic decision-making and generative AI in medicine have focused on bias and fairness. Robustness \cite{tocchetti2025} is an equally important topic, which generally refers to the consistency of model prediction to distribution shifts.
\begin{figure}[hbt!]
\centering
\includegraphics[width=0.55\textwidth]{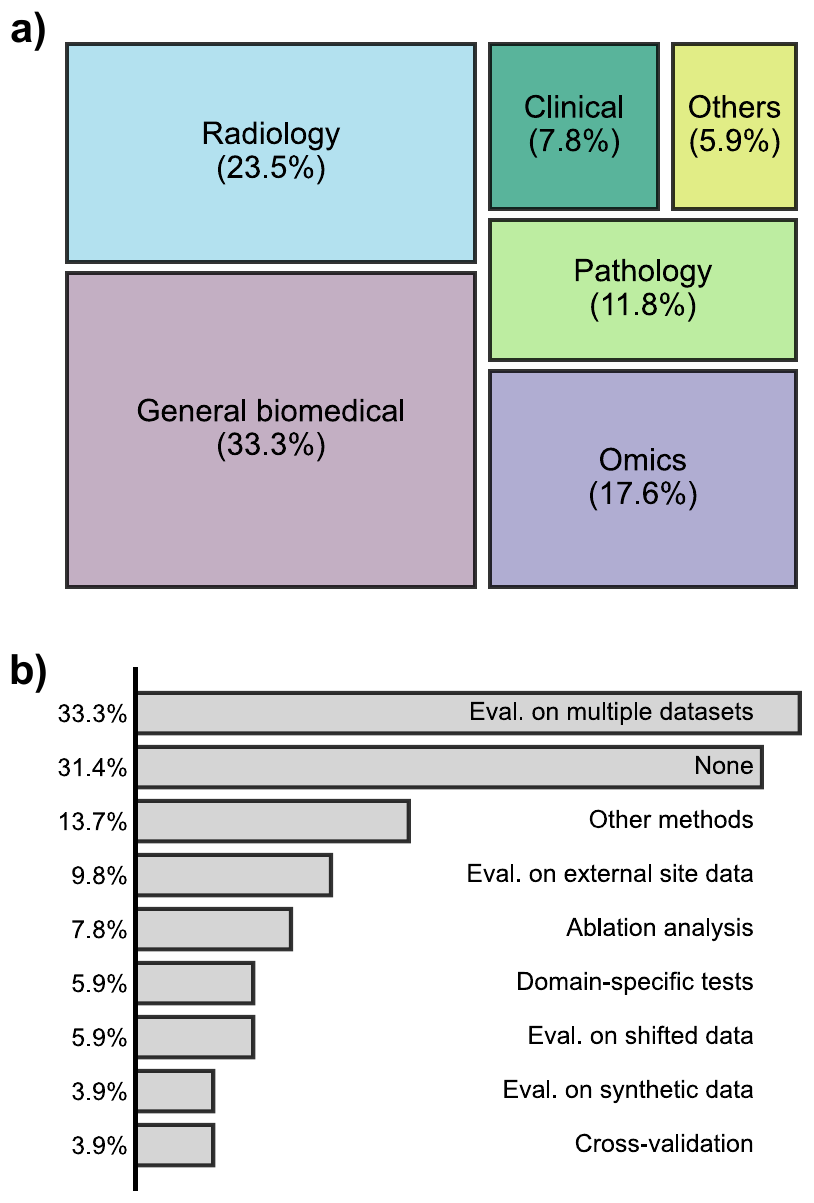}
\vspace{4pt}
\caption{\textbf{Existing robustness tests used for biomedical foundation models.} The treemap in \textbf{a} illustrates the topical areas of the BFMs looked at for this study. ``General biomedical'' indicates that the model is trained on general-purpose biomedical datasets and no domain specialization is emphasized in the model description. \textbf{b} shows the distributions of robustness tests (eval. = evaluation). Because multiple tests were conducted for some models, the total proportion in \textbf{b} is larger than unity.}
\label{fig:robtests}
\vspace{-0.5em}
\end{figure}
It is quantified using aggregated performance metrics, stratified comparisons across subsets of data, and worst-case performance. Robustness failures are an origin of the performance gap between model development and deployment, performance degradation over time, and, more alarmingly, the generation of misleading or harmful content by imperfect users or bad actors \cite{imperfect_users_2022}. The robustness of software also affects the legal responsibilities of providers \cite{Ladkin2020} because the software may cause harm (e.g. misinformation, financial loss, or injury) to users or third parties or require regulatory body authorization for deployment (e.g. medical devices) \cite{warraich_fda_2025,freyer_future_2024}.

\begin{figure*}[hbt!]
\centering
\includegraphics[width=0.95\textwidth]{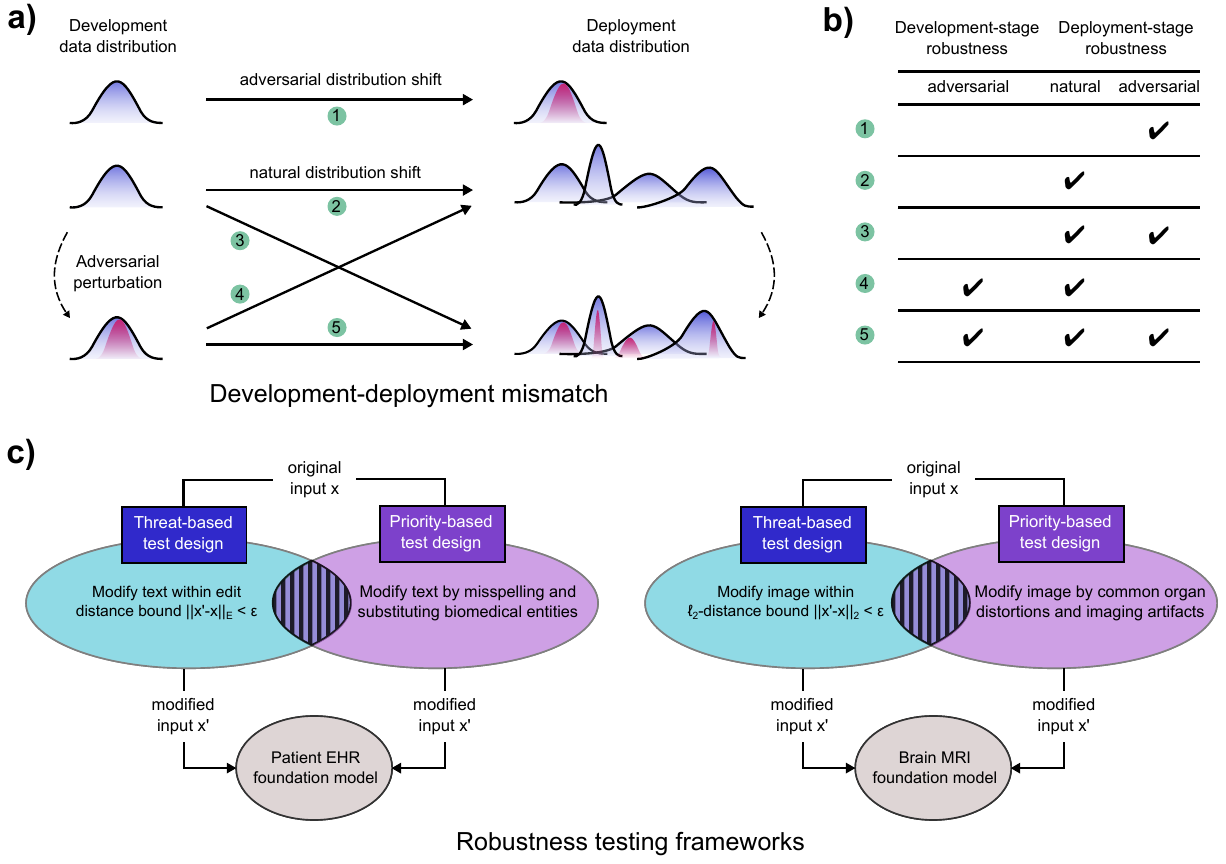}
\vspace{4pt}
\caption{\textbf{Settings and designs of robustness tests.} Visualization in \textbf{a} illustrates the potential settings of development and deployment mismatches, which are represented in \textbf{b} according to the types of distribution shifts. Setting 1 indicates adversarial contribution shift. Setting 2 refers to the natural distribution shift. In setting 3, adversarial perturbations are introduced in deployment, while in setting 4, they are applied to the training data. Setting 5 contains adversarial perturbations both during model development and deployment, such as in backdoor attacks. \textbf{c}, Specification of robustness by a simplified threat model (defined by a distance bound) or priority (defined by realistic artifacts) in the task domain, shown with two examples. The threat-based robustness tests use the error bound from edit distance for the EHR foundation model (left) and the Euclidean distance for the MRI foundation model (right). An overlap exists between these two approaches to generating test examples.}
\label{fig:cases}
\vspace{-0.5em}
\end{figure*}

We examined over 50 existing BFMs covering different biomedical domains (see Fig. \ref{fig:robtests} and Supplementary Data 1). About 31.4\% of them contain no robustness assessments at all. The most commonly presented evidence of model robustness is consistent performance across multiple datasets, which is adopted in 33.3\% of the selected BFMs. Despite being a convenient proxy, consistent performance is not equivalent to a rigorous robustness guarantee because the relationships between datasets are generally unknown. Evaluations on shifted (5.9\%) or synthetic data (3.9\%), or data from external sites (9.8\%) can be more effective but are not yet popular. To ensure the constructive and beneficial use of BFMs, we need to consider robustness evaluation across the model lifecycle and in intended application settings \cite{lyell_more_2023}. In biomedical domains, the various robustness concepts that warrant consideration (see Box 1 and the \href{https://github.com/RealPolitiX/bfm-robust}{repo}) motivate test customization. Inspired by test case prioritization in software engineering \citep{rothermel_prioritizing_2001}, which improves the cost-effectiveness of software testing by focusing on important test scenarios, we suggest designing effective robustness tests according to task-dependent robustness specifications constructed from priority scenarios (or priorities, see Fig. \ref{fig:cases}c) to facilitate test standardization while utilizing existing specialized tests as building blocks. Next, we introduce our proposal along with the background and motivation.

\section*{The robustness evaluation challenges}
\noindent\textbf{Foundation model characteristics}. The versatility of use cases and exposure to complex distribution shifts are two major challenges of robustness evaluation (or testing) \cite{chen_foundational_2022} for foundation models that differentiate from prior generations of predictive algorithms. The versatility comes from foundation models' increased capabilities at inference time with knowledge injection through in-context learning, instruction following, the use of external tools (e.g. function calling) and data sources (e.g. retrieval augmentation), and with user steering of model behavior using specially designed prompts. These new learning paradigms blur the line between development and deployment stages and open up more avenues where models are exploited for their design imperfections.
\begin{figure*}[hbt!]
\vspace{-0.5em}
\begin{tcolorbox}[colback=Gray!40, colframe=Black, left=.5mm, right=.5mm, top=.5mm, bottom=.5mm, sharp corners, label=concepts, segmentation style=solid, equal height group=bottombox]
\setstretch{1.05}
\textbf{\large{Box 1}: Varieties of robustness concepts}\tcbline
\vspace{-10pt}
\setlength\columnsep{2pt}
\begin{multicols}{2}
\begin{tcolorbox}[colback=White, colframe=Gray!40, left=1mm, right=1mm, top=1mm, bottom=1mm, sharp corners, title=Theoretical perspective, coltitle=Black, fonttitle=\bfseries]
\small{
\textbf{Natural robustness} refers to the resistance to natural variations of data distribution, also called non-adversarial robustness to emphasize the distinction.
\vspace{3pt}\\
\textbf{Adversarial robustness} is a model's resistance to accidental or deliberate malicious data manipulations such as adversarial attacks and jailbreaks.
\vspace{3pt}\\
\textbf{Interventional robustness} refers to the invariance of model behavior to causally-defined interventions (e.g. image background or contrast shift, patient age shift) that characterize distribution shifts.
\vspace{3pt}\\
\textbf{Out-of-distribution robustness} is a model's resistance to semantic or covariate shift of test examples from the training data distribution.
\vspace{3pt}\\
\textbf{Average-case \& Worst-case robustness} define the robustness concepts based on either the average or worst-case performance metric.
\vspace{3pt}\\
\textbf{Feature \& Data robustness} define the robustness concepts based on the model performance change by removing or adding specific features or data in model development.
}
\end{tcolorbox}
\begin{tcolorbox}[colback=white, colframe=Gray!40, left=1mm, right=1mm, top=1mm, bottom=1mm, sharp corners, title=Engineering perspective, coltitle=Black, fonttitle=\bfseries]
\small{
\textbf{Development-stage robustness} resists corruption during model development by noise in labels or data, or by adversarial attacks with poisoning and backdoor attacks as the most common examples.
\vspace{3pt}\\
\textbf{Deployment-stage robustness} resists corruption at inference time due to transformed inputs, distracting information, or ill-intentioned behavior manipulations such as prompt injection and jailbreaks.
\vspace{3pt}\\
\textbf{Component \& Aggregated robustness} refer to the
}
\end{tcolorbox}
\begin{tcolorbox}[colback=white, colframe=Gray!40, left=1mm, right=1mm, top=1mm, bottom=1mm, sharp corners, coltitle=Black, fonttitle=\bfseries]
\small{
individual-level robustness or combined robustness of the addressable components in a multi-expert or multiagent system.
}
\end{tcolorbox}
\begin{tcolorbox}[colback=white, colframe=Gray!40, left=1mm, right=1mm, top=1mm, bottom=1mm, sharp corners, title=Application perspective, coltitle=Black, fonttitle=\bfseries]
\small{
\textbf{Group robustness} refers to the resistance of model performance against subpopulation distribution shifts. It is also called subgroup robustness.
\vspace{3pt}\\
\textbf{Instance robustness} refers to the consistency of model performance to label-preserving changes at the level of a particular or the worst-case instance.
\vspace{0.2em}\\
\textbf{Longitudinal robustness} refers to the consistency of model performance against temporal shifts or nonstationary effects in the data distribution.
\vspace{3pt}\\
\textbf{Knowledge robustness} refers to the consistency of model knowledge against perturbations at the level of entities and their relations, against missing or nonstationary elements in a knowledge graph used for development or inference.
\vspace{3pt}\\
\textbf{Vendor robustness} refers to the consistency of model prediction to instrument vendors or the associated acquisition protocols that generate data. It is also known as acquisition-shift robustness.
\vspace{3pt}\\
\textbf{Uncertainty-aware robustness} defines a model's ability to account for uncertainties in the data distribution or in the model's internal mechanisms.
\vspace{3pt}\\
\textbf{Prevalence-shift robustness} refers to the consistency of model performance to changes in disease prevalence from epidemiological transitions.
\vspace{3pt}\\
\textbf{Behavioral robustness} refers to the consistency of model behavior in continuous interactions with humans or other models in a changing environment.
}
\end{tcolorbox}
\end{multicols}
\end{tcolorbox}
\vspace{-0.8em}
\end{figure*}

Distribution shifts arise from natural changes in the data or intentional and sometimes malicious data manipulation (i.e. adversarial distribution shift) \cite{chen_foundational_2022}. However, their distinction is increasingly nuanced in the era of foundation models \cite{qi_safe_secure_2024} due to the growing number of use cases. Natural distribution shifts can manifest biomedically in changing disease symptomatology, divergent population structure, etc. Inadvertent text deletion or image cropping results in data manipulations, potentially leading to adversarial examples that alter model behavior.

More elaborate shifts have been designed by targeted manipulation in model development and deployment \cite{yang_poisoning_2024,jin_backdoor_2024} through the cybersecurity lens. Poisoning attacks involve stealthy modification of training data, while in backdoor attacks, a specific token sequence (called a trigger) is inserted during model training and activated during inference time \cite{chowdhury_breaking_2024}. Distribution shifts in the deployment stage result in the majority of failure modes, including input transforms applied to text (deletion, substitution, and addition, including prompt injection, jailbreaks, etc) or images (noising, rotation, cropping, etc). Both natural distribution shifts and data manipulation yield out-of-distribution data \cite{karunanayake_ood_2025}. They can have high domain-specificity or be created to target specific aspects of the model lifecycle, resulting in complex origins that are hard to trace exactly.

\vspace{1em}\noindent\textbf{Robustness framework limitations}. Aside from the challenges in scope, how to generate appropriate test examples for robustness evaluation is not often discussed. Two important robustness frameworks in ML, adversarial and interventional robustness, come from the security and causality viewpoints, respectively. The adversarial framework typically requires a guided search of test examples within a distance-bounded constraint, such as the bounds established by edit distance for text and by Euclidean distance for image in Fig. \ref{fig:cases}c, yet there is no practical guarantee that the test examples are sufficiently naturalistic to reflect reality. The interventional framework requires predefined interventions and a corresponding causal graph, which is not immediately available for every task. Theoretical guarantees provided by these frameworks generally require justifications in the asymptotic limit and don't necessarily translate into effective robustness in diverse yet highly contextualized deployment settings of specialized domains \cite{hager_evaluation_2024,johri_evaluation_2025}. Because robustness testing (and hence its associated guarantee) is critically dependent on the robustness framework of choice, we should design robustness tests that are more aligned with naturalistic settings and reflective of the priorities in corresponding domains.

\section*{Specifying robustness by priorities}
\hspace*{1.5em}Effective robustness evaluations require a pragmatic framework. The two aspects central to its specification are: (i) the degradation mechanism behind a distribution shift, and (ii) the task performance metric that requires protection against the shift. Mechanistically understanding a robustness failure mode requires establishing a connection between (i) and (ii), which is costly when accounting for every type of user interaction or impractical when the users have insufficient information on model development history or blackbox access. Moreover, multiple degradation mechanisms can simultaneously affect a particular downstream task.

Technical robustness evaluations in ML have generally tackled simplified threats for obtaining statistical guarantees, where a specific degradation mechanism guides the creation of test examples. Most adversarial and interventional robustness tests fit into this category \cite{qi_safe_secure_2024}, which often targets a considerably broader set of scenarios than those that are meaningful in reality. From the efficiency perspective, taking a priority-based viewpoint \cite{rothermel_prioritizing_2001} and focusing on retaining task performance under commonly anticipated degradation mechanisms in deployment settings is sufficient. Robustness tests based on simplified threat models and priorities are not mutually exclusive because accounting for realistic and meaningful perturbations (priority-based) has certain overlap with distance-bounded perturbations (threat-based), while the outcomes of priority-based tests should directly inform model quality. Fig. \ref{fig:cases}c contains two examples comparing threat- and priority-based robustness tests for text and image data inputs. It illustrates the relationship between these two approaches for designing robustness tests.
\begin{figure*}[hbtp!]
\begin{tcolorbox}[colback=Gray!40, colframe=Black, left=.5mm, right=.5mm, top=.5mm, bottom=.5mm, sharp corners, label=specs, segmentation style=solid]
\setlength\columnsep{2pt}
\setstretch{1.05}
\textbf{\large{Box 2}: Examples of robustness specifications}\tcbline
\vspace{-10pt}
\begin{multicols}{2}
\begin{tcolorbox}[colback=White, colframe=Gray!40, left=1mm, right=1mm, top=1mm, bottom=1mm, sharp corners, title=Pharmacy chatbot for OTC medication, coltitle=Black, fonttitle=\bfseries]
\small{
The chatbot which can generate a medication request for patients has been extensively tested for robustness under the following scenarios:
\begin{enumerate}[wide, labelindent=0pt]
\item Up to 12 turns of conversation on topics related to primary care medication
\item Knowledge of specialized vocabulary, including products of major pharmaceutical vendors
\item Partly missing patient information (may refuse the task if patient is unwilling to provide sufficient information)
\item Drug combinations, recommended intake quantities, and dosage limits
\item Adverse drug interactions and patient history
\item Stock availability and provider preference
\item Limited prescription authority (will refuse the task if asked for non-OTC or prescription drugs)
\item Paraphrases and typos in the instruction prompt
\item Off-topic requests and non-existent drugs
\end{enumerate}
}
\end{tcolorbox}
\begin{tcolorbox}[colback=White, colframe=Gray!40, left=1mm, right=1mm, top=1mm, bottom=1mm, sharp corners, title=MRI radiology report copilot, coltitle=Black, fonttitle=\bfseries]
\small{
The copilot that generates a report from one or a series of images has been extensively tested for robustness under the following scenarios:
\begin{enumerate}[wide, labelindent=0pt]
\item Up to 15 turns of conversation including comparison of up to five images
\item Data from top-four vendors of MRI scanners
\item Variability of MRI acquisition mode, image contrast, cropping or masking, and resolution
\item Common imaging artifacts (may refuse the task if image quality is too low)
\item Major human organs and body sections (may refuse the task if images given don't resemble human anatomy)
\item Knowledge of specialized vocabulary
\item Spatial relations of human anatomy
\item Temporal ordering of distinguishing disease manifestations
\item Paraphrases and typos in the instruction prompt
\item Off-topic requests and non-MRI images
\end{enumerate}
}
\end{tcolorbox}
\end{multicols}
\end{tcolorbox}
\end{figure*}

We refer to the collection of priorities that demand testing for an individual task as a \textit{robustness specification}. To contextualize it in naturalistic settings, we constructed two examples in Box 2 for an LLM-based pharmacy chatbot for over-the-counter (OTC) medicine and a VLM-based radiology report copilot for magnetic resonance imaging (MRI), both of which are attainable with existing research in BFM development. The specification contains a mixture of domain-specific (e.g. drug interaction, scanner information) and general aspects (e.g. paraphrasing, off-topic requests) that can induce model failures. The specification breaks down robustness evaluation into operationalizable units such that each is convertible into a small number of quantitative tests with guarantees. In reality, the test examples may come from augmenting or modifying the specified information in an existing data record \cite{hager_evaluation_2024,johri_evaluation_2025}, such as a clinical vignette or case report. The specification can accommodate the future capability expansion of models and risk assessment updates accordingly. We discuss below the feasibility of our proposal using existing and potential realizations of major types of robustness tests for BFMs in application settings (see Box 1).

\vspace{1em}\noindent\textbf{Knowledge integrity.} BFMs are knowledge models and the knowledge acquisition process in the model lifecycle can be tempered to compromise knowledge robustness. Demonstrated examples for BFMs include a poisoning attack on biomedical entities, which have been shown to affect an entire knowledge graph in LLM-based biomedical reasoning \citep{yang_poisoning_2024} and a backdoor attack using noise as the trigger for model failures in MedCLIP \cite{jin_backdoor_2024}. Testing knowledge robustness should focus on knowledge integrity checks using realistic transforms. For text inputs, one may prioritize typos and distracting domain-specific information involving biomedical entities over random string perturbation under an edit-distance limit (see Fig. \ref{fig:cases}b). Existing examples include deliberately misinforming the model about the patient history \cite{han_medical_2024}, negating scientific findings \cite{yan_worse_2025}, and substituting biomedical entities \cite{xian_assessing_2024} to induce erroneous model behaviors. For image inputs, one may prioritize the effects of common imaging and scanner artifacts \cite{boone_rood-mri_2023}, alterations in organ morphology and orientation on model performance (see Fig. \ref{fig:cases}b).

\vspace{1em}\noindent\textbf{Population structure.} Explicit or implicit group structures are often present in biomedical and healthcare data, including prominent examples such as subpopulations organized by age group, ethnicity, or socioeconomic strata, medical study cohorts with specific phenotypic traits, etc. BFM-enabled cross-sectional or longitudinal studies for patient similarity analysis and health trajectory simulation may lead to group or longitudinal robustness issues when evaluating on incompatible populations. Group robustness assesses the model performance gap between the best- and worst-performing groups, either identifiable through the label or hidden in the dataset. Testing group robustness may modify subpopulation labels in patient descriptions to gauge the change in model performance \cite{yang_subpop_2023}. At a finer granularity, instance robustness represents the performance gap between instances that are more prone to robustness failures than others, which are likely corner cases. It is important when the model deployment setting requires a minimum robustness threshold for every instance. Robustness testing in this context may use a balanced metric to reflect the impact of input modifications across individual instances.

\vspace{1em}\noindent\textbf{Uncertainty awareness.} The machine learning community typically distinguishes between aleatoric uncertainty, which comes from inherent data variability, and epistemic uncertainty, which arises from insufficient knowledge of the model in the specific problem context. Robustness tests against aleatoric uncertainty may assess the sensitivity of model output to prompt formatting and paraphrasing, while assessing robustness to epistemic uncertainty may use out-of-context examples \cite{chandu2025} to examine if a model acknowledges the significant missing contextual information in domain-specific cases (e.g. presenting the model with a chest X-ray image and asking for a knee injury diagnosis). Additionally, uncertain information may also be directly verbalized in text prompts, a fitting scenario in biomedicine, to examine its influence on model behavior. Overall, the current generation of robustness evaluations hasn't yet included realistic uncertain scenarios often encountered in medical decision-making, although robustness against uncertainty is an important topic in practice.

\section*{Embracing emerging complexities}
\hspace*{1.5em}Previous scenarios primarily consider assessing a monolithic model using single-criterion robustness tests. Specifying and testing robustness for more complex AI systems should also account for performance tradeoffs, model architecture, and user interactions.
\begin{table*}[htbp]
\setstretch{1.5}
    \centering
    \caption{Robustness tests in the adaptation and update of BFM-based devices and services.}
    \vspace{0.5em}
    \begin{tabular}{|c|c|c|c|}
         \hline
         \textbf{Adaptation / Update} & \textbf{Execution} & \textbf{Cost} & \textbf{Potential usage of robustness tests} \\
         \hline
         \multirow{3}{*}{\parbox{2.2cm}{\setstretch{1}\centering Model weights}} & \multirow{3}{*}{\parbox{2.2cm}{\setstretch{1}\centering Model provider}} & \multirow{3}{*}{\parbox{1.8cm}{\setstretch{1}\centering high}} & \multirow{3}{*}{\parbox{7cm}{\setstretch{1}Helps choose and optimize the training method that yields models with a better robustness-accuracy tradeoff}} \\
         & & & \\
         & & & \\
         \hline
         \multirow{3}{*}{\parbox{2.2cm}{\setstretch{1}\centering Model architecture}} & \multirow{3}{*}{\parbox{2.2cm}{\setstretch{1}\centering Model provider}} & \multirow{3}{*}{\parbox{1.8cm}{\setstretch{1}\centering high}} & \multirow{3}{*}{\parbox{7cm}{\setstretch{1}Helps choose architectural parameters that are easier to attain a robust model than others when training on the same dataset}}\\
         & & & \\
         & & & \\
         \hline
         \multirow{2}{*}{\parbox{2.2cm}{\setstretch{1}\centering User prompt}} & \multirow{2}{*}{\parbox{2.2cm}{\setstretch{1}\centering Model user}} & \multirow{2}{*}{\parbox{1.8cm}{\setstretch{1}\centering low}} & \multirow{2}{*}{\parbox{7cm}{\setstretch{1}Helps select more robust prompt templates and improve the reliability of outputs}} \\
         & & & \\
         \hline
         \multirow{2}{*}{\parbox{2.2cm}{\setstretch{1}\centering User confidence}} & \multirow{2}{*}{\parbox{2.2cm}{\centering Model user}} & \multirow{2}{*}{\parbox{1.8cm}{\setstretch{1}\centering medium}} & \multirow{2}{*}{\parbox{7cm}{\setstretch{1} Helps adjust and calibrate user confidence to deploy the model for appropriate tasks}} \\
         & & & \\
         \hline
    \end{tabular}
    \label{tab:test_benefits}
    \vspace{-1em}
\end{table*}

\vspace{1em}\noindent\textbf{Metrics and stakeholders.} Evaluating tradeoffs between various robustness metrics and criteria offers a balanced view of a model's robustness across different dimensions and through metric aggregation. These more comprehensive robustness tests are essential in assessing whether the model's behavior reaches an optimal balance or is suitable for applications with distinct risk levels or stakeholders (see Supplementary Information section 1). When models are integrated into a healthcare workflow, they can affect downstream biomedical outcomes. For example, using LLMs to summarize or VLMs to generate case reports may influence clinician decisions by emphasizing certain conditions or sentiments, affecting diagnoses or procedures. This highlights the need for considering robustness tests with the relevant stakeholder(s) in the loop and behavioral robustness across diverse interaction settings to assess the model's impact on the care journey.

\vspace{1em}\noindent\textbf{Compound systems.} As modularity and maintainability become increasingly important, decision-making will be delegated to specialized subunits in a multi-expert (such as a mixture of experts) or multiagent system \citep{wang_masurvey_2025} with a centralized coordinating unit. In these compound AI systems, each addressable subsystem is subject to testing and maintenance according to capability demand and regulatory compliance (see Fig. \ref{fig:cases}c). For example, Polaris \cite{mukherjee_polaris_2024} from Hippocratic AI features a multiagent medical foundation model that writes medical reports and notes as well as engages in low-risk patient interactions. Future systems with specialized units can mimic the group decision-making process in healthcare \cite{radcliffe_collective_2019} to manage real-world complexities through enhanced reasoning and cooperative performance gain. Robustness tests for compound AI systems may consider different specifications for subsystems depending on the part-part and part-whole relationship in identifying bottlenecks and cascading effects associated with robustness failures.
\vspace{-0.2em}

\section*{Bridging policy with implementation}
\hspace*{1.5em}Ensuring robustness for BFMs requires advancing regulatory policies for both AI and health information technology. Currently, the leading AI regulatory frameworks, such as the EU AI Act and the US Federal AI Risk Management Act, recognize the relation between natural and adversarial notions of robustness but contain insufficient details to guide implementation in domain-specific applications (see Supplementary Information section 2). Existing health information technology regulations, such as the US-based HTI-1 final rule by the Office of the National Coordinator, focus primarily on transparency and disclosures of the use of predictive decision support models, yet lack detailed robustness requirements. The situation is in part due to the lack of a safety bare minimum for specific biomedical applications \citep{freyer_future_2024} and the fast-evolving technological landscape, which can exacerbate the challenges laid out at the beginning of this Comment. These existing gaps make concrete community-endorsed standards on robustness even more important.

\vspace{1em}\noindent\textbf{Considerations in implementation.} Mandating robustness specifications according to the tasks and the biomedical domains (see Fig. \ref{fig:robtests}-\ref{fig:cases}) provides a means to map policy objectives onto real-world implementations. It also facilitates evidence collection and enables effective risk management throughout the model lifecycle. We advocate that robustness specifications (i) should seek community endorsement to gain a broad adoption; (ii) should consider the permissible tasks and user group characteristics due to the difference in user journeys; (iii) should inform regulatory standards such as in the construction of quantitative risk thresholds \cite{koessler_risk_2024} or safety cases by enriching the failure mode taxonomy of BFMs and improving their informativeness. These considerations will facilitate the implementation of robustness specifications and ensure that their adoption is within shared interests of stakeholders.

\vspace{1em}\noindent\textbf{Community benefits.} Establishing a consensus-driven robustness specification from the research community will incentivize systematic efforts by model developers, research institutions, and independent third parties. For model developers, robustness testing informs model selection and updates. For the model deployment team and model users, robustness testing allows for identifying inference-time adjustments of prompt templates to improve the reliability of BFM applications. These potential uses of robustness tests are summarized in Table \ref{tab:test_benefits}. In addition, robustness specifications provide templates for failure-reporting procedures to allow users to provide timely feedback to the deployment team. Integrating robustness specifications with incident reporting mechanisms \cite{lyell_more_2023} facilitates the identification of model vulnerabilities and guides targeted improvements or informs post-hoc adjustments to model behavior. Their implementations can assist the training of end-users to recognize potential failures, calibrate user confidence, and enact mitigation strategies.

\section*{Acknowledgements}
R.P.X. thanks N. Rethmeier for helpful discussions. R.A.-A. would like to acknowledge funding from the Weill Neurohub. M.S. is partially funded by the National Cancer Institute of the National Institutes of Health under Award Number P30CA082103. The content is solely the responsibility of the authors and does not necessarily represent the official views of the National Institutes of Health.

\section*{Author contributions}
R.P.X. and R.A.-A. conceptualized the main idea of the work. R.P.X. prepared the figures and tables, and wrote the majority of the text with significant contributions from N.R.B., T.D., Q.C., and A.J.H. S.B. provided theoretical support. M.S. and R.A.-A. provided background information in different application settings. All authors edited the text and contributed to the discussions to finalize the manuscript.

\section*{Competing interests}
T.D. is a co-founder and director of governance \& standardization at PRISM Eval. Other authors declare no competing interests.

\section*{Additional information}
The extended resource on the various concepts of robustness mentioned in Box 1 is available from an \href{https://github.com/RealPolitiX/bfm-robust}{online repository}. The data on biomedical foundation models used for Fig. \ref{fig:robtests} are available as Supplementary Data 1.

\section*{Supplementary Information}
\section{Application-specific robustness metrics}

The construction of robustness metrics is specific to the use cases because of the data modality and practical requirements involved. Among the three types of robustness metrics, aggregated metrics allow a balanced view of robustness failures and are used as a general assessment. Stratified comparisons across distinct subgroups (e.g. demographics, clinical contexts, temporal shifts, or biological characteristics) offer a comprehensive evaluation of both model performance and ethical alignment. Worst-case metrics set a lower bound on the model performance and are more useful in high-risk settings where the negative effects should be considered fully. In the following, we consider three commonly encountered use cases in biomedical applications and discuss the ways to construct the relevant metrics:

In \textbf{diagnostic decision support}, the demographic information is usually directly taken into account. The most common performance metric is accuracy \citep{miller_diagnostic_2016}. Evaluations for robustness should include comparison across distinct subgroups such as those defined by age, sex, and race of the patient. The robustness metrics for this task should take into account the disparity between subgroups or use the worst-subgroup accuracy across stratified demographic subgroups.

In \textbf{medical image interpretation} or \textbf{medical report generation}, the common metrics include semantic overlap such as ROUGE \citep{lin_rouge_2004}, BERTScore \citep{zhang_bertscore_2019}, and more specialized variants like RadGraph F1 \citep{yu_evaluating_2023}, which accounts for relationship and completeness at the level of biomedical named entities in the generated interpretation or report. While medical images do not explicitly encode race information, they do reveal key biological characteristics such as organ morphology, anatomical variation, and biological age and sex. Constructing robustness metrics should account for the model’s aggregated performance using out-of-distribution data that include the effects of common image distortion and shifts along biologically relevant covariates that represent anatomical variations.

In \textbf{clinical text summarization}, demographic information is often present due to the nature of the text data. The common performance metrics in summarization are semantic overlap and faithfulness (aka. factual correctness). Semantic overlap is like just discussed for the previous task. Quantifying faithfulness requires extraction of the factual component from both the original text and the summarization before comparison \citep{maynez_faithfulness_2020}. Robustness evaluations for this task should consider text dataset shifts including typos, grammatical errors, variations in narrative style, and documentation practices across institutions. The robust metrics for this task can be an aggregated metric or the difference between stratified subgroups bearing their sensitive subgroup information.


\section{Robustness in major AI regulatory frameworks}

We provide here more details on the robustness requirements in AI systems from major AI policy recommendations and regulations in the European Union (EU) and United States (US). We quote the corresponding documents wherever needed to illustrate the details presented there.\vspace{0.5em}

The \textbf{EU AI Act} is the first regulation of AI by a major jurisdiction, the EU. It considers robustness and cybersecurity as related concepts and puts AI models in biomedical and healthcare applications within the high-risk AI systems category \citep{bellogin_eu_2024}. The AI Act delineates various requirements on accuracy, robustness and cybersecurity together \citep{nolte_robustness_2025} in its Article 15 (\url{https://artificialintelligenceact.eu/article/15/}), which will go into force in August 2026. Regarding natural robustness, the AI Act demands that high-risk AI systems ``shall be as resilient as possible regarding errors, faults or inconsistencies that may occur within the system or the environment in which the system operates, in particular due to their interaction with natural persons or other systems.'' Regarding adversarial robustness (considered within the scope of cybersecurity in the AI Act), the AI Act demands that high-risk AI systems ``shall be resilient against attempts by unauthorised third parties to alter their use, outputs or performance by exploiting system vulnerabilities.''
\vspace{1em}

The \textbf{US Federal AI Risk Management Act} promotes the AI risk management framework (\url{https://www.nist.gov/itl/ai-risk-management-framework}) and its successors developed by the US National Institute of Standards and Technology (NIST). It is currently a leading framework on the subject issued by a US federal agency \citep{rawal_responsible_2024}, but it is yet to be enacted into law as of mid-2025 (\url{https://www.govinfo.gov/app/details/BILLS-118hr6936ih/}). The NIST framework is a set of recommendations and it identifies the measurement of risk, tolerance determination of risk, and prioritization of risks as the major challenges in AI risk management. The NIST framework adopts an industry- and use case-agnostic approach and refers to resilience as the counterpart of robustness that also accounts for the resistance to ``adversarial use of model or data''. The NIST framework encompasses four key processes: map, measure, manage, and govern to be implemented throughout the lifecycle of AI systems. Robustness is featured in the measure process, where the framework mentions that ``The AI system to be deployed is demonstrated to be safe, its residual negative risk does not exceed the risk tolerance, and it can fail safely, particularly if made to operate beyond its knowledge limits. Safety metrics reflect system reliability and robustness, real-time monitoring, and response times for AI system failures.''

\section*{Supplementary data 1}
\textbf{Collected data on robustness tests for biomedical foundation models.} The data from over 50 biomedical foundation models include information on the model developers (e.g. institutions), the modality (e.g. language, vision, or both), model capabilities, biomedical domain, and types of robustness tests, along with reference to the respective publication. They are used for creating Figure 1 in the main text.

\setstretch{1.3}
\bibliography{sn-bibliography}

\end{document}